\pdfoutput=1
\documentclass[reprint,aps,prx,showpacs,twocolumn]{revtex4-1}

\usepackage{natbib}
\usepackage{graphicx}
\usepackage[utf8]{inputenc}
\usepackage{mathptmx}
\usepackage{epsfig,amsopn}
\usepackage{graphicx}
\usepackage{amsmath,amssymb}
\usepackage{natbib,color}
\usepackage{braket}
\usepackage{float}
\usepackage{enumerate}
\usepackage[normalem]{ulem}
\usepackage{todonotes}
\allowdisplaybreaks[1]

\newcommand{\eref}[1]{Eq.~(\ref{#1})}%
\newcommand{\fref}[1]{Fig.~\ref{#1}} %
\newcommand{\sgn}[1]{\mathrm{sgn}({#1})}%
\newcommand{\erfc}{\mathrm{erfc}}%
\newcommand{\erf}{\mathrm{erf}}%

\def\bea{\begin{eqnarray}}
\def\eea{\end{eqnarray}}

\DeclareMathOperator{\sign}{sgn}

\begin{document}

\title{Diffusion with stochastic resetting is invariant to return speed}

\author{Arnab Pal$^{1}$}
\email{arnabpal@mail.tau.ac.il}

\author{\L{}ukasz Ku\'smierz$^{2}$}
\email{nalewkoz@gmail.com}

\author{Shlomi Reuveni$^{1}$}
\email{shlomire@tauex.tau.ac.il}

\affiliation{\noindent \textit{$^{1}$School of Chemistry, The Center for Physics and Chemistry of Living Systems, The Raymond and Beverly Sackler Center for Computational Molecular and Materials Science,\\ \& The Mark Ratner Institute for Single Molecule Chemistry, Tel Aviv University, Tel Aviv 6997801, Israel}}

\affiliation{\noindent \textit{$^{2}$Laboratory for Neural Computation and Adaptation, RIKEN Center for Brain Science, 2-1 Hirosawa, Wako, Saitama 351-0198, Japan}}

\date{\today}

\begin{abstract}
The canonical Evans--Majumdar model for diffusion with stochastic resetting to the origin assumes that resetting takes zero time: upon resetting the diffusing particle is teleported back to the origin to start its motion anew. However, in reality getting from one place to another takes a finite amount of time which must be accounted for as diffusion with resetting already serves as a model for a myriad of processes in physics and beyond. Here we consider a situation where upon resetting the diffusing particle returns to the origin at a finite (rather than infinite) speed. This creates a coupling between the particle's random position at the moment of resetting and its return time, and further gives rise to a non-trivial cross-talk between two separate phases of motion: the diffusive phase and the return phase. We show that each of these phases relaxes to the stead-state in a unique manner; and while this could have also rendered the total relaxation dynamics extremely non-trivial our analysis surprisingly reveals otherwise. Indeed, the time-dependent distribution describing the particle's position in our model is completely invariant to the speed of return. Thus, whether returns are slow or fast, we always recover the result originally obtained for diffusion with instantaneous returns to the origin.  

\end{abstract}

\maketitle

\section{Introduction}
Diffusion with stochastic resetting has drawn considerable attention in recent times due to its rich non-equilibrium properties and various applications to first-passage theory. Consider a diffusive particle that, at random moments in time, is stopped and instantaneously returned to its initial position \cite{Restart1,Restart2}. Recent studies have shown that such a procedure can have dramatic consequences on the underlying process. For example, in contrast to free diffusion, the resetting system attains a non-equilibrium steady state at long times \cite{Restart1,Restart2,KM,restart_conc3}; and is moreover characterized by non-trivial relaxation dynamics. The propagator has an inner core region near the resetting point which relaxes to the steady-state while the outer region continues to evolve with time; The border separating the two regions grows with time according to a power law \cite{restart_conc5}. 

The simple model of diffusion with stochastic resetting has  been extended and generalized to cover: diffusion in the presence of a potential \cite{restart_conc2,Ray,Ahmad}, in a domain \cite{Christou,restart_conc8,localtimer,restart_conc9}, and arbitrary dimensions \cite{restart_conc6}; diffusion in the presence of non-exponential resetting time distributions e.g., deterministic \cite{Restart-Search3}, intermittent \cite{restart_conc3}, non-Markovian \cite{restart_conc18}, 
non-stationary \cite{kusmierz2018robust}, with general time dependent resetting rates \cite{Restart4}, as well as other protocols \cite{Restart5}; and diffusion in the presence of interactions \cite{restart_conc1,restart_conc12,SEP}. The effect of resetting on random walks \cite{restart_conc7,restart_conc16}, continuous time random walks \cite{Restart3,restart_conc21,subCTRW},  L\'evy flights \cite{Restart-Search1,Restart-Search2}, and other forms of stochastic motion \cite{Satya-RT,telegraphic,transport1,Bodrova1,Bodrova2}, has also been studied.

One of the most salient effects of resetting on a diffusing particle is that it renders its mean first-passage time to a target finite \cite{Restart1,Restart2}. The observation that resetting can also expedite the completion of other stochastic processes has thereafter opened an active research front focused on first-passage under restart \cite{PalReuveniPRL,Chechkin,ReuveniPRL,branching,branching_I} with immediate applications to physical chemistry and biological physics \cite{Restart-Biophysics1,Restart-Biophysics2,Restart-Biophysics3,Restart-Biophysics4,Restart-Biophysics5,Restart-Biophysics6,Restart-Biophysics7,Restart-Biophysics8}. The effects of generic resetting protocols on arbitrary first-passage processes were characterized \cite{PalReuveniPRL}, sharp (deterministic) restart was proven to be a winning restart strategy which cannot be beaten in terms of mean first-passage time minimization \cite{PalReuveniPRL}, and additional universalities associated with first-passage under restart were also found \cite{ReuveniPRL,branching,Landau}. 

\begin{figure}[t]
\centering
\includegraphics[width=\linewidth]{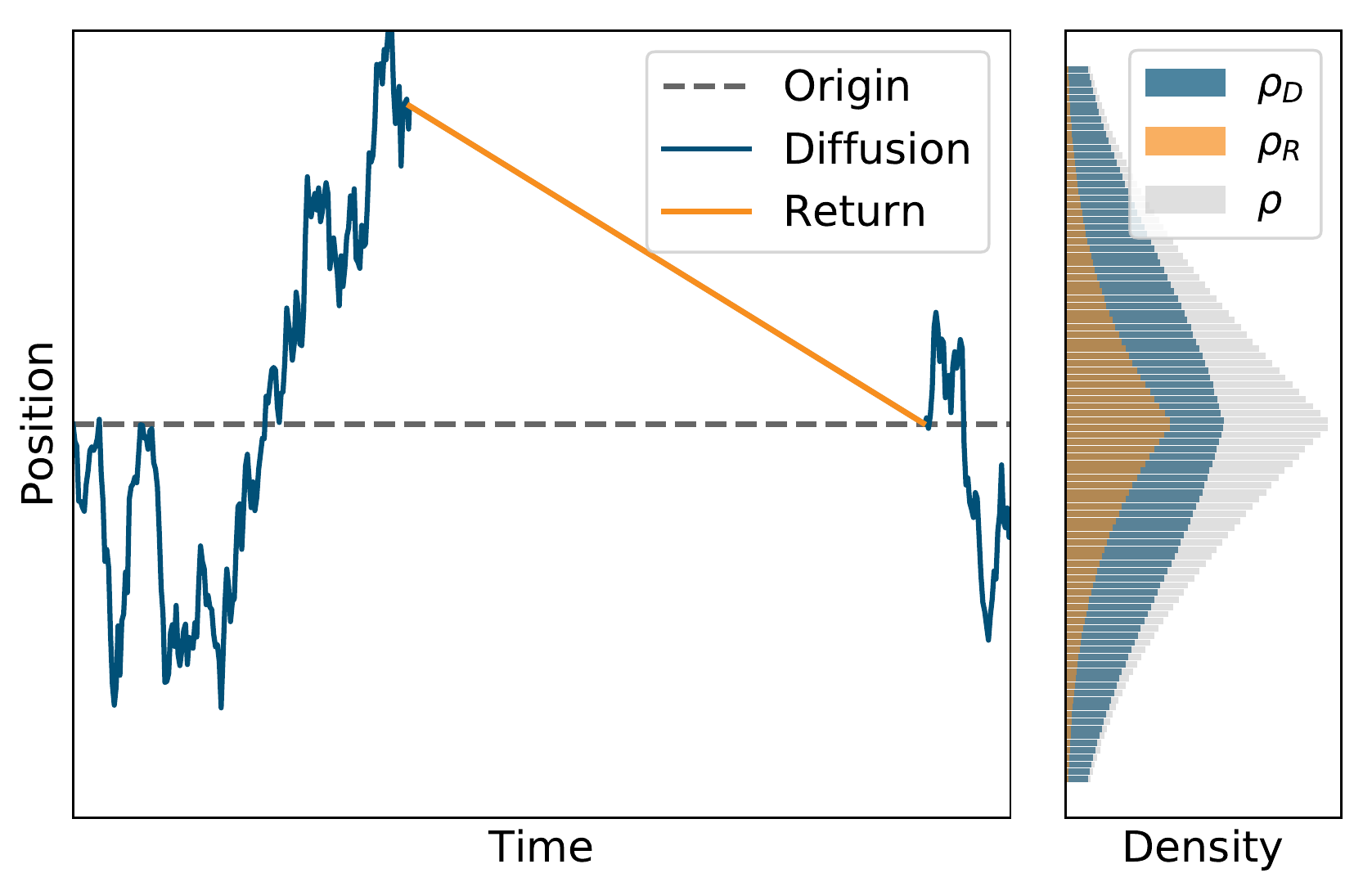}
\caption{An illustration of diffusion with stochastic resetting and constant speed returns to origin. The dynamics consists of two modes: (i) diffusion (blue), wherein the particle moves diffusively; and (ii) return (orange), wherein upon resetting the particle returns to the origin at a constant speed. Here we study the steady-state and relaxation dynamics of the probability densities $\rho_D$ and $\rho_R$ which govern the diffusive and return phases respectively. We find that the steady-state and relaxation dynamics of the overall density $\rho=\rho_D+\rho_R$ is completely invariant to the speed at which the particle returns to the origin.}
\label{fig:illustration}
\end{figure}

The basic model of diffusion under resetting and most of its variants considered resetting to be instantaneous i.e., upon resetting the particle is stopped and instantaneously returned to its initial position. However, in reality it is clear that a particle cannot return (or be returned) to the origin in zero time. Several attempts were then made to address this issue e.g., by incorporating an overhead time (refractory period) that follows each resetting event \cite{Restart-Biophysics1,Restart-Biophysics2,Ahmad,transport1,Satya-refractory}; but in all these attempts it was assumed that there is no direct coupling between the overhead time and the position of the particle at the resetting moment---which is clearly non-physical since returning from afar usually takes longer. To address this issue, we have recently introduced a comprehensive theory for first-passage under space-time coupled resetting, a.k.a, home-range search, which does not make any assumptions on the underlying stochastic motion and is furthermore suited to treat generic return and home-stay strategies \cite{HRS}. In this complementary study, we set aside first-passage questions and focus on the relaxation and steady-state properties of simple diffusion with stochastic resetting and constant speed returns to the origin (Fig.~\ref{fig:illustration}). As we will shortly demonstrate, this intriguing case study exhibits a surprising invariance with respect to the return speed. 

\section{Diffusion with stochastic resetting and constant speed returns}
\label{Sec2}

Consider a Brownian particle which starts its motion at the origin at time zero and performs simple diffusion. The motion of such a particle is described by the diffusion equation
\begin{equation}
    \partial_t \rho(x,t) = 
    D \partial_x^2 \rho(x,t),
    \label{eq:simple_diff}
\end{equation}
where $D$ is the diffusion constant and $\rho(x,t)$ is the propagator, i.e., the probability density to find the particle at position $x$ at time $t$. To introduce  stochastic resetting with instantaneous returns into the model imagine that at any small time interval $\Delta t$ the particle's motion can be reset with probability $r\Delta t$. If such resetting happens, the particle will teleport back to the origin and start its motion anew. The corresponding master equation then reads  \cite{Restart1,Restart2}
\bea
\partial_t\rho(x,t)=D\partial_x^2 \rho(x,t)-r \rho(x,t)+r \delta(x)~.
\label{inst-propagator}
\eea
In this section, we will construct and solve a set of master equations, akin to \eref{inst-propagator}, that describe diffusion with stochastic resetting and non-instantaneous returns. In particular, we will consider a situation where returns are conducted at a constant speed $v_r$ which means that the time taken to return from position $x$ is simply $|x|/v_r$. 

Similar to the above, we will once again denote the propagator by $\rho(x,t)$, but we will now discriminate between two different phases of motion: (i) the diffusive phase in which the particle performs Brownian motion; and (ii) the return phase in which the particle returns to its initial position. Our propagator thus has two contributions, one from each phase, and it can be written as \bea
\rho(x,t)=\rho_D(x,t)+\rho_R(x,t)~,
\eea
where $\rho_D(x,t)$ and $\rho_R(x,t)$ correspond to the probability densities of finding the particle in the diffusion and return phases respectively. It is clear that $\rho_D(x,t)$ and $\rho_R(x,t)$ are not individually normalized as their sum is the total probability density $\rho(x,t)$ which is normalized to one. Evidently the probabilities to find the particle in the diffusion and return phases are given by
\begin{align}
\begin{split}
    p_D(t) \equiv \text{Prob}(\text{\textit{diffusion}}) &= \int\limits_{-\infty}^{\infty} \mathrm{d}x\rho_D(x,t), 
    \\
    p_R(t) \equiv \text{Prob}(\text{\textit{return}}) &= \int\limits_{-\infty}^{\infty} \mathrm{d}x\rho_R(x,t),
    \label{pa-pb}
\end{split}
\end{align}
where $p_D(t)+p_R(t)=1$ at all times. 

We now set to find the propagator $\rho(x,t)$ which describes our process. We start by considering the time evolution of the position distribution in the return phase $\rho_R(x,t)$. To this end, we recall that particles in the return phase move at a speed $v_r$ in the direction of the origin, i.e., to the left if $x>0$ and to the right if $x<0$. The probability flux at $x$ due to such particles is thus $\partial_x[\sgn{x}v_r\rho_R(x,t)]$ where $\sgn{x}$ stands for the signum function which takes the value: $+1$ if $x>0$, $-1$ if $x<0$, and zero otherwise. In addition, we note that particles enter the return phase from the diffusion phase at a rate $r$. The probability flux at $x$ due to such particles is thus $r\rho_D(x,t)$. Summing over the two possibilities above gives 
\bea
    \partial_t \rho_R(x,t) &=&
    \partial_x[\sgn{x}v_r\rho_R(x,t)]+ r\rho_D(x,t) \nonumber
    \\
     \nonumber
    \\
    &-&
    2\delta(x)v_r\rho_R(x,t), 
    \label{return-delta}
\eea
where the last term on the right hand side serves as a sink and accounts for the fact that returning particles switch to diffusive mode upon arrival to the origin. Finally, we observe that taking the spatial derivative on the right side of \eref{return-delta} cancels the last term and leaves us with  
\begin{equation}
    \partial_t \rho_R(x,t) = 
    \sign{(x)} v_r \partial_x  \rho_R(x,t) + r \rho_D(x,t)~. 
    \label{eq:balance-dynamics-zombie-main}
\end{equation}

We now turn our attention to the time evolution of the position distribution in the diffusive phase $\rho_D(x,t)$. Proceeding as before, we observe that a diffusing particle will be found at position $x$ at time $t+\Delta t$ if at time $t$ it was positioned at $x-\Delta x$ and provided that in the following time interval $\Delta t$ it diffused an increment $\Delta x$. Noting that the probability to continue diffusing within this latter time interval is $(1-r\Delta t)$ and setting $\Delta x=\sqrt{2D\Delta t}\xi$ to be the corresponding infinitesimal displacement ($\xi$ is a standard normal random variable), we have 
\bea
\rho_D(x,t+\Delta t) &=& (1-r\Delta t) \langle \rho_D(x-\Delta x,t)   \rangle \nonumber \\ &+&\delta(x)~\int\limits_{-v_r \Delta t}^{ v_r \Delta t}~dz~\rho_R(z,t) 
+ O(\Delta t^2),
\eea
where the average in the first term is taken with respect to the random variable $\xi$, and the second term acts as a source which accounts for the inflow of diffusing particles at the origin due to particles returning from the domain $\left[ -v_r \Delta t,v_r \Delta t  \right]$ and consequently switching to diffusive mode. Taking the $\Delta t\to 0$ limit, the corresponding continuous-time evolution equation reads
\begin{equation}
    \partial_t \rho_D(x,t) = D\partial_x^2  \rho_D(x,t)-r \rho_D(x,t)
    +2 v_r \rho_R(0,t) \delta(x)~,
    \label{eq:balance-dynamics-active-main}
\end{equation}
where we assume, without much loss of generality, 
that $\rho_R(x,t)$ is also continuous around ${x=0}$. 

Equations (\ref{eq:balance-dynamics-zombie-main}) and (\ref{eq:balance-dynamics-active-main}) constitute a set of two coupled partial differential equations that can be solved in the Laplace space. Following Laplace transform, these equations read
\bea
\sign(x)v_r \partial_x \tilde{\rho}_R(x,s)-s\tilde{\rho}_R(x,s)+r\tilde{\rho}_D(x,s)&=&0~,\hspace{0.65cm}
\label{Aeq1} \\
D\partial_{x}^2 \tilde{\rho}_D(x,s)-(r+s)\tilde{\rho}_D(x,s)+\left[ 2v_r \tilde{\rho}_R(0,s)+1 \right] \delta(x)&=&0~,\hspace{0.65cm}
\label{Aeq2}
\eea
where we have defined $\tilde{\rho}_R(x,s)=\int_0^\infty~dt~e^{-st} \rho_R(x,t)$, $\tilde{\rho}_D(x,s)=\int_0^\infty~dt~e^{-st} \rho_D(x,t)$, and further assumed that the particle has started in the diffusion phase at time zero i.e., $\rho_D(x,0)=\delta(x)$.
Solving \eref{Aeq2} with the natural boundary conditions $\underset{x \to \pm \infty}{\mathrm{lim}} \tilde{\rho}_D(x,s)=0$, we obtain
\bea
\tilde{\rho}_D(x,s)=A ~\exp \left[  -\sqrt{\frac{r+s}{D}}~ |x| \right]~,
\label{Aeq3}
\eea
where 
\bea
A=\frac{2v_r\tilde{\rho}_R(0,s)+1}{\sqrt{4D(r+s)}}~.
\label{Aeq3_A}
\eea
Substituting \eref{Aeq3} into \eref{Aeq1}, we obtain
\bea
 \tilde{\rho}_R(x,s)=\frac{Ar}{s+\sqrt{\frac{r+s}{D}}~v_r}~\exp \left[  -\sqrt{\frac{r+s}{D}} ~|x| \right]~.
 \label{Aeq4}
\eea

To find an explicit expression for the constant $A$ in \eref{Aeq3_A}, we first need to compute $\tilde{\rho}_R(0,s)$. This is done by setting $x=0$ in \eref{Aeq4} and utilizing \eref{Aeq3_A} to solve for $\tilde{\rho}_R(0,s)$. This gives
\bea
\tilde{\rho}_R(0,s)=\frac{r}{2s}\frac{1}{v_r+\sqrt{D(r+s)}}~.
 \label{Aeq5}
\eea
Substituting \eref{Aeq5} into the expression for $A$ in \eref{Aeq3_A} gives 
\bea
A=\frac{1}{2s}\frac{s+v_r\sqrt{(r+s)/D}}{v_r+\sqrt{D(r+s)}}~.
 \label{Aeq6}
\eea
Substituting \eref{Aeq6} into \eref{Aeq3} and \eref{Aeq4}, we conclude that
\begin{align}
    \begin{split}
    \tilde{\rho}_R(x,s) &= 
    \frac{1}{2 s} \frac{r}{v_r + \sqrt{D(r+s)}} 
    e^{-\sqrt{(r+s)/{D}}~|x|}, 
    \\
    \tilde{\rho}_D(x,s) &= 
    \frac{1}{2 s} \frac{s + v_r \sqrt{(r+s)/D}}{v_r + \sqrt{D(r+s)}} 
    e^{-\sqrt{(r+s)/{D}}~|x|}~.
    \end{split}
    \label{LT-jointPDF}
\end{align}

\section{Diffusion with stochastic resetting is invariant to speed of return to origin}\label{Sec3}

It is clear from \eref{LT-jointPDF} that the probability densities describing the diffusing and returning phases depend on the return speed $v_r$, as expected. It is, however, particularly striking to observe that the total density is given by 
\bea
    \tilde{\rho}(x,s) = \tilde{\rho}_R(x,s)+ \tilde{\rho}_D(x,s)=
    \frac{1}{2 s} 
    \sqrt{\frac{r+s}{D}}
    e^{-\sqrt{(r+s)/{D}}|x|},\hspace{0.4cm}
    \label{LT-propagator-slow}
\eea
which means that it is completely invariant to the return speed. Moreover, since the model of diffusion with stochastic resetting and instantaneous returns can be obtained from our model by taking the limit $v_r \to \infty$, the density in \eref{LT-propagator-slow} must be identical to that which was obtained in \cite{Restart1,Restart2} for this limiting case. 

To see the invariance result more directly recall that the propagator of diffusion with stochastic resetting and  instantaneous returns can be linked to that of simple diffusion by taking advantage of a renewal approach \cite{Restart1,Restart2}. In Laplace space, this connection reads
\bea
\tilde{\rho}_{\mathrm{\infty}}(x,s)=\frac{r+s}{s}~\tilde{\rho}_{SD}(x,r+s)
\label{renewal-general}
\eea
where $\tilde{\rho}_{\mathrm{\infty}}(x,s)$ and   $\tilde{\rho}_{SD}(x,s)$ are the Laplace space solutions of  \eref{inst-propagator} (our model in the $v_r \to \infty$ limit) and \eref{eq:simple_diff} (simple diffusion) respectively. Since the propagator of simple diffusion is Gaussian, we have $\tilde{\rho}_{SD}(x,s)=\int_0^\infty~dt~e^{-st} e^{-x^2/4Dt}/\sqrt{4\pi D t}=e^{-\sqrt{s/D}|x|}/\sqrt{(4Ds)}$, which by use of  \eref{renewal-general} leads to 
\bea
 \tilde{\rho}_{\mathrm{\infty}}(x,s) = 
    \frac{1}{2 s} 
    \sqrt{\frac{r+s}{D}}
    e^{-\sqrt{(r+s)/{D}}|x|}~.
    \label{equivalence}
\eea
The right hand side of \eref{equivalence} is identical to the right hand side of  \eref{LT-propagator-slow}, which means that the propagator of diffusion with stochastic resetting and instantaneous returns is identical to the propagator of diffusion with stochastic resetting and constant velocity returns. Thus, for all times and irrespective of the return velocity we have
\bea
\rho(x,t) = \rho_{\mathrm{\infty}}(x,t).
\label{equivalence2}
\eea
To delve deeper and explicitly show this invariance and additional properties of our processes, we now invert Eqs. (\ref{LT-jointPDF}) and (\ref{LT-propagator-slow}) to present real time solutions for $\rho_D(x,t)$, $\rho_R(x,t)$, and $\rho(x,t)$.  

\section{Transient and steady-state solutions}\label{Sec4}

Inverting \eref{LT-propagator-slow}, we compute the propagator governing diffusion with stochastic resetting and constant speed returns and find that it is given by
\bea
\rho(x,t)&=&\mathcal{L}^{-1}\left[     \frac{1}{2 s} 
    \sqrt{\frac{r+s}{D}}
    e^{-\sqrt{(r+s)/{D}}|x|} \right] \nonumber\\
    &=&f(x)+b(x,t)~, \label{Total_Real}
\eea
where $f(x)=\frac{\alpha_0}{2}e^{-\alpha_0|x|}$ with $\alpha_0=\sqrt{\frac{r}{D}}$ standing for the inverse mean distance traveled by the particle while it is in the diffusive phase \cite{Restart1}, and with 
\bea
b(x,t)&=&e^{-rt}~\frac{\exp\big[-\frac{x^2}{4Dt}\big]}{\sqrt{4\pi Dt}}-\frac{\alpha_0}{2}\cosh\big[\alpha_0 x\big]\nonumber \\
&+&\frac{\alpha_0}{4}~\exp\big[\alpha_0 x \big]~\text{erf}\bigg[\frac{x+2Dt\alpha_0}{\sqrt{4 Dt}}\bigg]\nonumber \\
&+&\frac{\alpha_0}{4}~\exp\big[-\alpha_0 x \big]~\text{erf}\bigg[\frac{-x+2Dt\alpha_0}{\sqrt{4 Dt}}\bigg].
\eea

The density of the returning phase is given by \eref{LT-jointPDF} which can be inverted using the method of complex inversion. This gives the density of the returning particles at all times in an integral form which can be easily computed numerically
\bea
\rho_R(x,t)&=& \mathcal{L}^{-1} \left[ \frac{1}{2 s} \frac{r}{v_r + \sqrt{D(r+s)}} 
    e^{-\sqrt{(r+s)/{D}}|x|} \right] \nonumber \\
&=&\int_0^t~dt'~g_1(t-t')~g_2(t')~,
\label{Returning_Real}
\eea
where 
\bea
g_1(t)=\frac{r e^{-r t}}{\sqrt{4\pi Dt}}+\sqrt{\frac{r^3}{4D}}~\erf \left( \sqrt{r t}\right)~,
\eea
and
\bea
g_2(t)= e^{-rt+v_r^2 t/D+v_r|x|/D} ~ \erfc \left( \frac{v_r \sqrt{t}}{\sqrt{D}}+\frac{|x|}{\sqrt{4Dt}} \right)~.
\eea
Since $\rho_D(x,t)=\rho(x,t)-\rho_R(x,t)$, the density of the diffusing particles can then be obtained by subtracting the result in \eref{Returning_Real} from the result in \eref{Total_Real}. 

\begin{figure}[t]
    \centering
    \includegraphics[width=\linewidth]{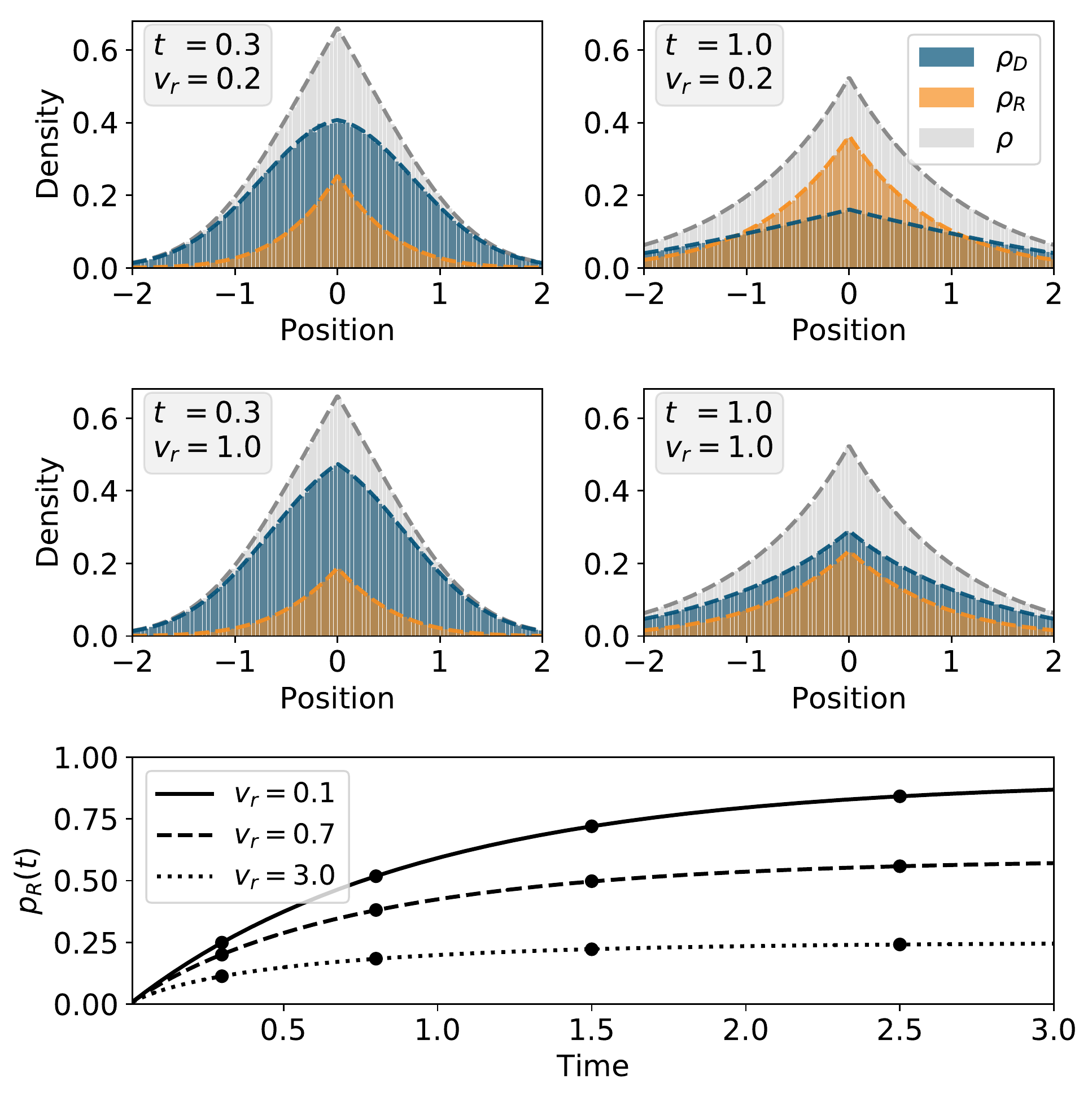}
    \caption{We simulated diffusion with stochastic resetting and compared the results with the analytical predictions of Eqs. (\ref{Total_Real})-(\ref{eq:pr-in-time}). 
    Top and middle: Densities at different times with $v_r = 0.2$ (top) and $v_r =1$ (middle). Histograms represent results of computer simulations, 
    dashed lines the corresponding analytical predictions.
    Bottom: The probability of being in the return phase as a function of time for different values of $v_r$. Results obtained via computer simulations 
    are represented by dots. 
    In all cases $10^6$ trajectories were generated with $r=D=1$ and $\Delta t = 10^{-3}$.
    }
    \label{fig:mc-check}
\end{figure}

An exact, real-time, formula can also be given for the probability to be in the return phase $p_R(t)$; and hence for the probability $p_D(t)=1-p_R(t)$ to be in the diffusive phase. Integrating \eref{Aeq4} over the space variable and inverting the Laplace transform we find
\begin{equation}
    p_R(t) = 
    \frac{r}{r_0 - r} \left(
        e^{(r_0 - r) t} \erfc\left(\sqrt{r_0 t}\right) +
        \sqrt{\frac{r_0}{r}} \erf\left(\sqrt{r t}\right) 
        - 1
    \right),
\label{eq:pr-in-time}
\end{equation}
where $r_0 = v^2_r/D$. 
Clearly, the relaxation is non-exponential. The results in Eqs. (\ref{Total_Real})-(\ref{eq:pr-in-time}) are successfully corroborated against numerical simulations in Fig.~\ref{fig:mc-check}.

\begin{figure*}[t]
    \centering
    \includegraphics[width=\linewidth,height=3.5cm]{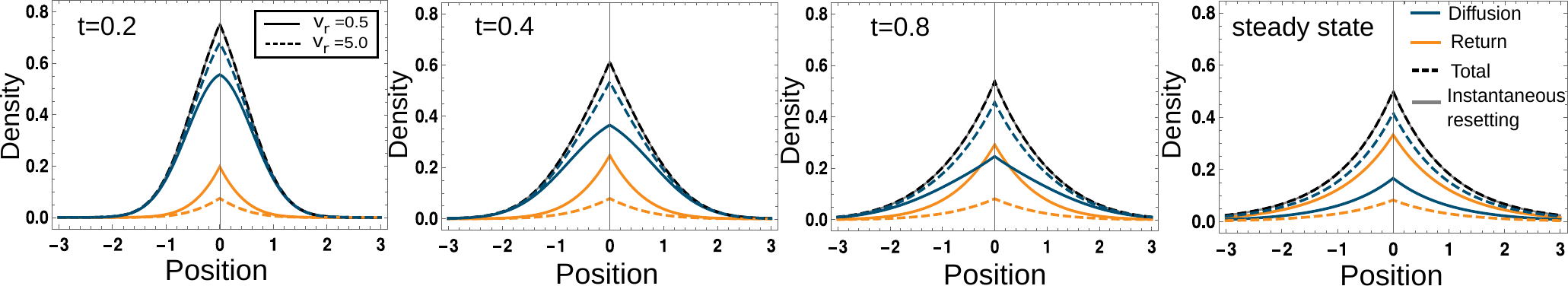}
    \caption{Illustration of the invariance properties of diffusion with stochastic resetting and constant speed returns to origin. The probability densities describing the diffusive phase $\rho_D(x,t)$ in blue, and the return phase $\rho_R(x,t)$ in orange, are plotted for different times $t=\{0.2,0.4,0.8,\infty$\}, and  different return speeds $v_r=0.5$ (solid lines) and $v_r=5.0$ (dashed lines). The total density $\rho(x,t)=\rho_D(x,t)+\rho_R(x,t)$ is also plotted (dashed black). One can observe that while the relaxation of the diffusive and return phases depend on the return speed the relaxation of the total density does not. Thus, as asserted by \eref{LT-propagator-slow}, this density remains the same regardless of whether returns are slow or fast. In particular, as asserted by \eref{equivalence2}, this means that the total density behaves exactly as it does when returns are instantaneous ($v_r\to\infty$ limit depicted in solid grey). Another interesting invariance is observed at the steady-state where all densities take on the same Laplace form (up to a scaling factor) as was shown in Eqs. (\ref{SS-diffusion-return1})-(\ref{SS-total}).}
    \label{fig:time-dep_invaariance}
\end{figure*}

The densities of the diffusive and returning phases in the steady state can be computed by taking $t\to \infty$ limit in their respective time-dependent expressions. But they can also easily be computed from \eref{LT-jointPDF} using the final value theorem. A simple exercise then gives
\bea
\rho_D(x)=\lim_{s\to 0} s ~\tilde{\rho}_D(x,s)&=&\frac{\alpha_0p_D}{2}e^{-\alpha_0|x|}
\label{SS-diffusion-return1}
~\\
\rho_R(x)=\lim_{s\to 0} s ~\tilde{\rho}_R(x,s)&=&\frac{\alpha_0p_R}{2} e^{-\alpha_0|x|}
\label{SS-diffusion-return2}
\eea
where we recall that $\alpha_0=\sqrt{\frac{r}{D}}$, and where 
$p_D= \frac{1/r}{1/r+\alpha_0^{-1}/v_r }$, and $p_R=1-p_D$ are the
steady-state probabilities to find the particle in the diffusive and return phase respectively. 

The results for the steady-state probabilities to find the particle in the different phases can be understood by observing that the particle spends $r^{-1}$ units of time on average in the diffusive phase and $\alpha_0^{-1}/{v_r}$ units of time on average in the return phase. Note, however, that while these probabilities depend on the return velocity $v_r$ they act only as scaling factors for the same Laplace, $\sim e^{-\alpha_0|x|}$, distribution. Thus, while the relaxation dynamics governing the diffusive and return phases are different, their steady-states are identical (up to a scaling factor). Moreover, in the steady state we have
\bea
\rho(x)=\rho_D(x)+\rho_R(x)=
\frac{\alpha_0}{2}e^{-\alpha_0|x|}
\label{SS-total}
\eea
which, as expected, is identical to the expression  found for the case of stochastic resetting with instantaneous returns \cite{Restart1}. This nicely demonstrates that the Laplace distribution governs all the steady-state distributions in our problem. 

\section{Conclusion}
\label{Sec5}

In this paper we studied diffusion with stochastic resetting. In contrast to previous models where resetting is followed by an instantaneous return to the origin, here we considered a more realistic scenario in which the particle returns to the origin at a constant speed. Surprisingly, we found that this generalization has no effect on the steady-state of the process and on its relaxation dynamics. Starting from the origin, the probability to find the particle at position $x$ at time $t$ is the same regardless of whether returns are slow or fast. Since instantaneous returns are recovered in the limit of infinitely high return speeds we find that the probability distribution in our model is identical to the one describing the original Evans–Majumdar model for diffusion with stochastic resetting and instantaneous returns. This result is demonstrated in \fref{fig:time-dep_invaariance} using the analytical expressions obtained in section \ref{Sec4}.

The finite return speed considered herein gives rise to two separate phases of motion: the return phase and the diffusive phase. The overall dynamics of the process is attained by summing over the two phases, but one could also study the two phases separately. We did so and found that while each phase has its own relaxation dynamics, which is furthermore different than the one governing the overall dynamics, the steady-state distribution of the two phases is identical (up to a scaling factor) and invariant to the return speed. Indeed, while the relative contribution of the diffusing and returning phases to the overall probability to find the particle at a given position strongly depends on the return speed, the steady-states themselves are governed by the same Laplace distribution that in turn also characterizes the steady-state distribution of the entire process. This result is also demonstrated in \fref{fig:time-dep_invaariance}.

Finally, we report that the invariance properties discovered here for the special case of diffusion could be traced back to a more fundamental invariance property which underlies stochastic processes with space-time coupled resetting. This is established and elaborately discussed in \cite{invariance}. 

\section{Author Contributions}
Arnab Pal and \L{}ukasz Ku\'smierz have contributed equally to this work.
\section{Acknowledgements}
All authors would like to acknowledge Tam\'{a}s Kiss, Sergey Denisov and Eli Barkai, organizers of the 672. WE-Heraeus Seminar: ``Search and Problem Solving by Random Walks'', as discussions that led to this work began there. Shlomi Reuveni would like to deeply acknowledge Sidney Redner for a series of joint discussions which led to this work. Shlomi Reuveni acknowledges support from the Azrieli Foundation and from the Raymond and Beverly Sackler Center for Computational Molecular and Materials Science at Tel Aviv University. Arnab Pal acknowledges support from the Raymond and Beverly Sackler Post-Doctoral Scholarship at Tel-Aviv University.

\end{document}